\documentclass[12pt,a4paper]{article}
\usepackage{graphicx}
\usepackage{times}
\newcounter{subfigure}

\title{\bf Mixing in magnetic OB stars}
\author{Thierry Morel\\
\vspace{1cm}\\
\normalsize Institute for Astrophysics and Geophysics, Li\`ege University, All\'ee du 6 Ao\^ut 17, 4000 Li\`ege,\\
\normalsize Belgium
}
\date{\mbox{}}
\begin{document}
\maketitle
\pagestyle{empty}
%
%
\def\bull{\vrule height .9ex width .8ex depth -.1ex}
\makeatletter
\def\ps@plain{\let\@mkboth\gobbletwo
\def\@oddhead{}\def\@oddfoot{\hfil\tiny\bull\quad
``The multi-wavelength view of hot, massive stars''; 39$^{\rm th}$ Li\`ege Int.\ Astroph.\ Coll., 12-16 July 2010 \quad\bull}%
\def\@evenhead{}\let\@evenfoot\@oddfoot}
\makeatother
%
%
\def\beginrefer{\section*{References}%
\begin{quotation}\mbox{}\par}
\def\refer#1\par{{\setlength{\parindent}{-\leftmargin}\indent#1\par}}
\def\endrefer{\end{quotation}}
%
%
{\noindent\small{\bf Abstract:} 
Recent observations have revealed the existence of a population of slowly-rotating, nitrogen-rich B dwarfs that are not predicted by evolutionary models including rotational mixing. However, as theoretical arguments suggest that magnetic processes may significantly increase the efficiency of the transport of the chemical elements, it is of importance to assess the extent of mixing in some known magnetic OB stars. We review our knowledge of the CNO abundance properties of these objects and present the first results of an NLTE abundance study of massive stars identified as being magnetic by the MiMeS collaboration. Although a nitrogen excess is often associated with the presence of a magnetic field, there is no evidence for a strict one-to-one correspondence between these two phenomena. This therefore suggests that other (still elusive) parameters may control the amount of mixing experienced by main-sequence OB stars.
}
%
%
\section{Context}
\begin{figure}[t]
\begin{minipage}{8cm}
\centering
\includegraphics[width=6.0cm]{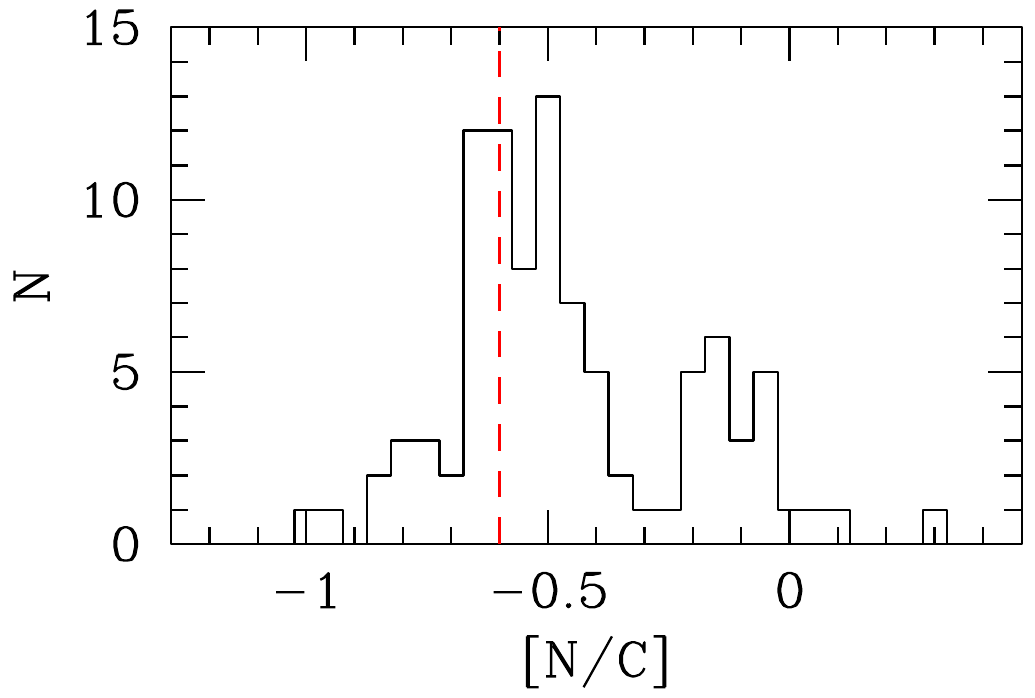}
\end{minipage}
\hfill
\begin{minipage}{8.5cm}
\caption{Distribution of the logarithmic N to C abundance ratio in nearby, main-sequence B stars (adapted from Morel 2009). The dashed line indicates the solar value (Asplund et al. 2009).\label{fig_histogram_nearby_CN}}
\end{minipage}
\end{figure}

Magnetic fields are involved in a wide variety of phenomena associated to massive stars. A question that has recently been a focus of interest is the impact they may have on mixing of the internal layers. Evolutionary models incorporating magnetic fields generated through dynamo action generally predict a greater amount of mixing and hence higher CNO abundance anomalies (Maeder \& Meynet 2005, but see Heger, Woosley \& Spruit 2005).  The detection of a sizeable population of slowly-rotating, yet N-rich, main-sequence B stars in the Magellanic clouds (Hunter et al. 2008) and in the Galaxy (e.g., Gies \& Lambert 1992; Kilian 1992; Morel, Hubrig \& Briquet 2008) challenges current rotational mixing theories (Brott et al. 2009) and urges the need to investigate this problem. The distribution of the N to C abundance ratio in nearby, main-sequence B stars appears to be bimodal with about 20--25\% exhibiting values a factor 2--3 higher than the bulk of the sample (Fig.\ref{fig_histogram_nearby_CN}). Guided by the theoretical results, one may be inclined to think that the relative proportion of magnetic stars could be higher in the N-rich group. Investigating the CNO abundance properties of magnetic OB stars thus appears not only warranted, but also timely in view of the rapidly growing number of magnetic field detections in massive stars. An indication for a higher incidence of a nitrogen excess in magnetic stars was inferred by Morel et al. (2008) who found 8 out of the 10 magnetic stars in their sample to be N rich by a factor $\sim$3. Here we re-address this result in the light of new spectropolarimetric observations that have questioned the magnetic status of some of these stars (Silvester et al. 2009) and first results of our NLTE abundance analyses of a number of main-sequence OB stars that have recently been shown to host a strong magnetic field by the MiMeS collaboration (Wade et al. 2010).

\section{Analysis of the MiMeS stars}
\subsection{Observations and targets}
High-resolution ($R$ $\sim$ 46,000) FIES spectra of four O9--B2 IV--V targets (NGC 2244 \#201, Par 1772, NU Ori and HD 57682) were obtained in late 2009 in the framework of the `fast-track service programme' of the Nordic Optical Telescope (NOT; Canary Islands). As can be seen in Fig.\ref{fig_spectra_targets}, these stars span a wide range of $v\sin i$ values (from about 22 to 225 km s$^{-1}$). 

Spectropolarimetric observations of NGC 2244 \#201 by Alecian et al. (2008) indicate a longitudinal field strength of about 500 G with no variability of the Stokes $V$ profile over 5 days. The star HD 57682 shows strong UV and optical line-profile variability. In particular, the H$\alpha$ profile is atypical and modulated by the rotational period (Grunhut et al. 2010). The relatively sharp emission component filling in the absorption profile observed in our data is very similar to the case reported by Halbedel (1993). The field likely confines the stellar wind and has a polar strength of $\sim$1700 G assuming a dipole morphology (Grunhut et al. 2009). The polar field strengths under the same assumption are $\sim$1150 and $\sim$600 G in Par 1772 and NU Ori, respectively (Petit et al. 2008).

\begin{figure}[h]
\centering
\includegraphics[width=10.3cm]{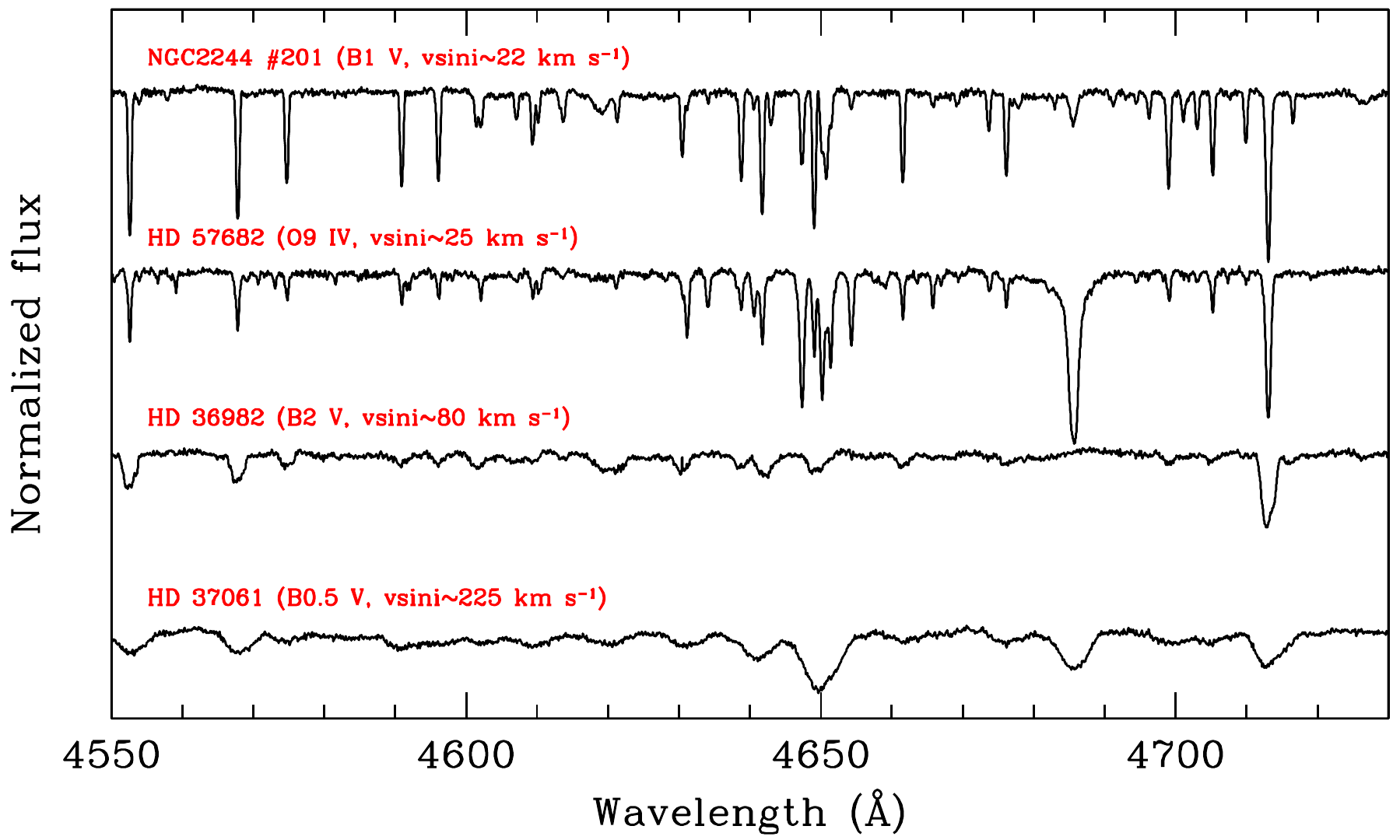}
\caption{Spectra of the targets for the spectral range 4550--4730 \AA.\label{fig_spectra_targets}}
\end{figure}

\subsection{Methods of analysis}
The atmospheric parameters are derived purely on spectroscopic grounds: $\log g$ is determined from fitting the collisionally-broadened wings of the Balmer lines, $T_{\rm eff}$ from ionisation balance of various species (He \,{\sc i/ii}, N \,{\sc ii/iii}, Ne \,{\sc i/ii} and/or Si \,{\sc iii/iv}) and the microturbulence, $\xi$, from requiring the abundances yielded by the O \,{\sc ii} features to be independent of their strength. The abundances are computed using Kurucz atmospheric models, an updated version of the NLTE line-formation codes DETAIL/SURFACE (Butler \& Giddings 1985; Giddings 1981) and classical curve-of-growth techniques. The relatively low mass-loss rate of the O9 IV star HD 57682 ($\dot{M}$ $\sim$ 1 $\times$ 10$^{-9}$ M$_{\odot}$ yr$^{-1}$; Grunhut et al. 2009) validates the use of a static model atmosphere for this star.
 
\subsection{First results}
In this contribution, we present the results for the two narrow-lined stars NGC 2244 \#201 and HD 57682. The two fast rotators remain to be analysed using spectral synthesis techniques. The atmospheric parameters and elemental abundances are provided in Table \ref{table_abundances} where they can be compared with previous results in the literature (Kilian 1992, 1994; Vrancken et al. 1997) and values obtained following exactly the same methodology for the magnetic, N-rich star $\tau$ Sco (Hubrig et al. 2008). Grunhut et al. (2009) obtained $T_{\rm eff}$=34500$\pm$1000 K and $\log g$=4.0$\pm$0.2 dex for HD 57682 from spectroscopic indicators using the NLTE, unified code CMFGEN (Hillier \& Miller 1998).

There is no indication for a contamination of the surface layers of these two main-sequence stars by core-processed material. In the case of NGC 2244 \#201, the CNO logarithmic abundance ratios ([N/C] and [N/O]) are consistent with the solar values, fully confirming the results of Vrancken et al. (1997). The results for HD 57682 are more uncertain owing to the weakness of the spectral lines and their strong $T_{\rm eff}$ sensitivity, but there is no indication for significant departures from the solar ratios either. Significant differences with the results of Kilian (1992, 1994) are found. 

\begin{table}
\begin{center}
\caption{Atmospheric parameters and elemental abundances of NGC 2244 \#201 and HD 57682 (on a scale in which $\log \epsilon$[H]=12). The results of previous studies in the literature (Kilian 1992, 1994; Vrancken et al. 1997) and those obtained for $\tau$ Sco using exactly the same tools and techniques are shown for comparison (Hubrig et al. 2008). The number of lines used is given in brackets. A blank indicates that no value could be determined. The solar [N/C] and [N/O] ratios are --0.60$\pm$0.08 and --0.86$\pm$0.08 dex, respectively (Asplund et al. 2009).}
\label{table_abundances}
\vspace*{0.3cm}
{\scriptsize
\begin{tabular}{l|cc|cc|c}\hline 
                        & \multicolumn{2}{c|}{NGC 2244 \#201}          & \multicolumn{2}{c|}{HD 57682}              & $\tau$ Sco\\
                        & This study          & Vrancken et al. (1997) & This study           & Kilian (1992, 1994) & Hubrig et al. (2008)\\\hline
$T_{\rm eff}$ (K)       & 27000$\pm$1000      & 27300$\pm$1000         & 33000$\pm$1000       & 31800$\pm$200       & 31500$\pm$1000\\
$\log g$ (cgs)          & 4.20$\pm$0.15       & 4.3$\pm$0.1            & 4.00$\pm$0.15        & 3.85$\pm$0.10       & 4.05$\pm$0.15\\
$\xi$ (km s$^{-1}$)     & 3$\pm$3             & 4                      & 5$\pm$5$^{a}$        & 0$^{+1}_{-0}$       & 2$\pm$2\\ 
$v\sin i$ (km s$^{-1}$) & 22$\pm$2            & 22$\pm$1.5             & 25$\pm$4             & 35$\pm$3            & 8$\pm$2\\
He/H                    & 0.072$\pm$0.023 (9) &                        & 0.106$\pm$0.030 (10) &  0.085$\pm$0.008    & 0.085$\pm$0.027 (9)\\
$\log \epsilon$(C)      & 8.22$\pm$0.13 (6)   & 8.20$\pm$ 0.23         & 8.20$\pm$0.19 (6)    &  8.75$\pm$0.06      & 8.19$\pm$0.14 (15)\\
$\log \epsilon$(N)      & 7.68$\pm$0.13 (20)  & 7.58$\pm$0.20          & 7.52$\pm$0.25 (8)    &  7.71$\pm$0.09      & 8.15$\pm$0.20 (35)\\
$\log \epsilon$(O)      & 8.63$\pm$0.18 (31)  & 8.59$\pm$0.19          & 8.31$\pm$0.21 (14)   &  8.10$\pm$0.08      & 8.62$\pm$0.20 (42)\\
$\log \epsilon$(Ne)     & 8.02$\pm$0.12 (7)   &                        & 7.95$\pm$0.17 (1)    &  8.11$\pm$0.06$^b$  & 7.97$\pm$0.10 (5)$^{c}$\\
$\log \epsilon$(Mg)     & 7.29$\pm$0.20 (1)   & 7.38                   & 7.37$\pm$0.18 (1)    &  7.33$\pm$0.07      & 7.45$\pm$0.09 (2)\\
$\log \epsilon$(Al)     & 6.20$\pm$0.13 (3)   & 6.15$\pm$0.15          & 6.07$\pm$0.21 (1)    &  6.23$\pm$0.06      & 6.31$\pm$0.29 (3)\\
$\log \epsilon$(Si)     & 7.41$\pm$0.25 (5)   & 7.28$\pm$0.30          & 7.47$\pm$0.32 (5)    &  7.24$\pm$0.06      & 7.24$\pm$0.14 (9)\\
$\log \epsilon$(S)      & 7.30$\pm$0.19 (1)   &                        &                      &  6.97$\pm$0.09$^b$  & 7.18$\pm$0.28 (3)\\ 
$\log \epsilon$(Fe)     & 7.33$\pm$0.13 (20)  &                        &                      &  7.48$\pm$0.12$^b$  & 7.33$\pm$0.31 (13)\\
${\rm [N/C]}$           & --0.54$\pm$0.14     & --0.62$\pm$0.31        & --0.68$\pm$0.30      & --1.04$\pm$0.11     & --0.04$\pm$0.25\\
${\rm [N/O]}$           & --0.95$\pm$0.21     & --1.01$\pm$0.28        & --0.79$\pm$0.19      & --0.39$\pm$0.13     & --0.47$\pm$0.29\\\hline
\end{tabular}
}
\end{center}
\scriptsize{$^a$: Assumed value. $^b$ LTE values. $^c$ From Morel \& Butler (2008).}
\end{table}

\section{Discussion}
These two apparently slowly-rotating, main-sequence stars do not show evidence for CN-cycled material at their surfaces and hence do not display the N excesses observed in other magnetic B stars. Figure \ref{fig_teff_logg1} shows the positions of these two stars in the $\log T_{\rm eff}$-$\log g$ plane, along with those of other late O/early B stars with or without a magnetic field detection analysed in exactly the same way (Morel et al. 2008). On average, stars in both groups have roughly similar masses, share about the same evolutionary status and are slow rotators. It is important to note that the distinction between magnetic and non-magnetic stars remains quite fuzzy: the detection in stars with weak fields is often disputed (see Hubrig et al. 2009 vs Silvester et al. 2009), while non magnetic stars might be detected with more sensitive and intensive observations. For this reason, we also show in Fig.\ref{fig_teff_logg2} the results for the stars for which the (non) detection of a magnetic field can be regarded as more secure, either because it is (un)detected at a high degree of confidence or because it is confirmed by independent studies. Although magnetic and abundance studies of a larger sample are needed to draw firm conclusions, in both cases there is evidence for a higher incidence of an N excess in the magnetic stars. 

\addtocounter{figure}{0}
\setcounter{subfigure}{1}
\begin{figure}[h]
\begin{minipage}{8cm}
\centering
\includegraphics[width=8cm]{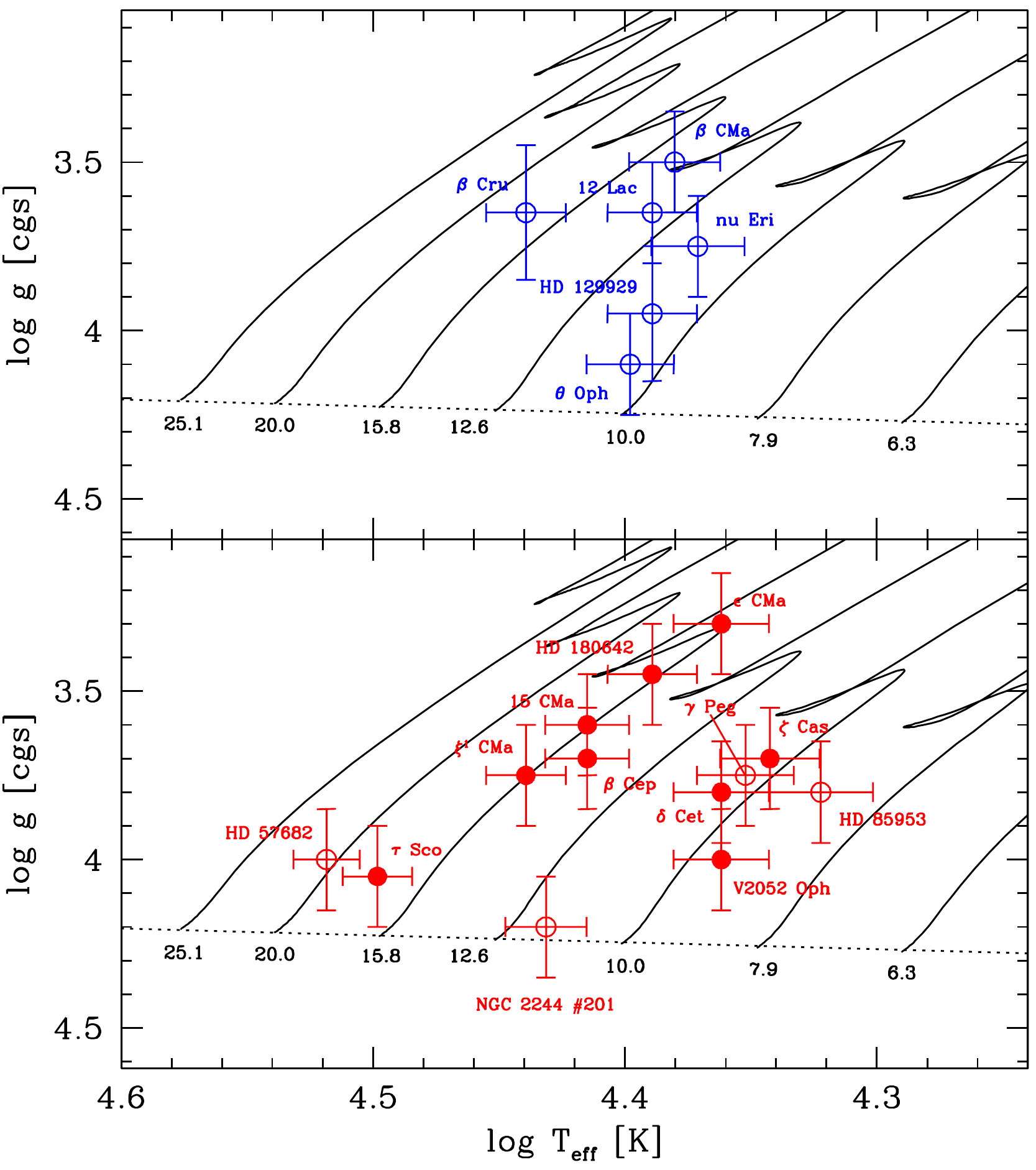}
\caption{Position in the $\log T_{\rm eff}$-$\log g$ plane of the OB stars without ({\it top panel}) and with ({\it bottom panel}) a magnetic field detection. Filled symbols denote stars showing an N excess (magnetic and abundance data from Morel et al. 2008 and this study). Evolutionary tracks from Claret (2004) for masses ranging from 6.3 to 25.1 M$_{\odot}$ are overplotted.\label{fig_teff_logg1}}
\end{minipage}
\hfill
\addtocounter{figure}{-1}
\addtocounter{subfigure}{1}
\begin{minipage}{8cm}
\centering
\includegraphics[width=8cm]{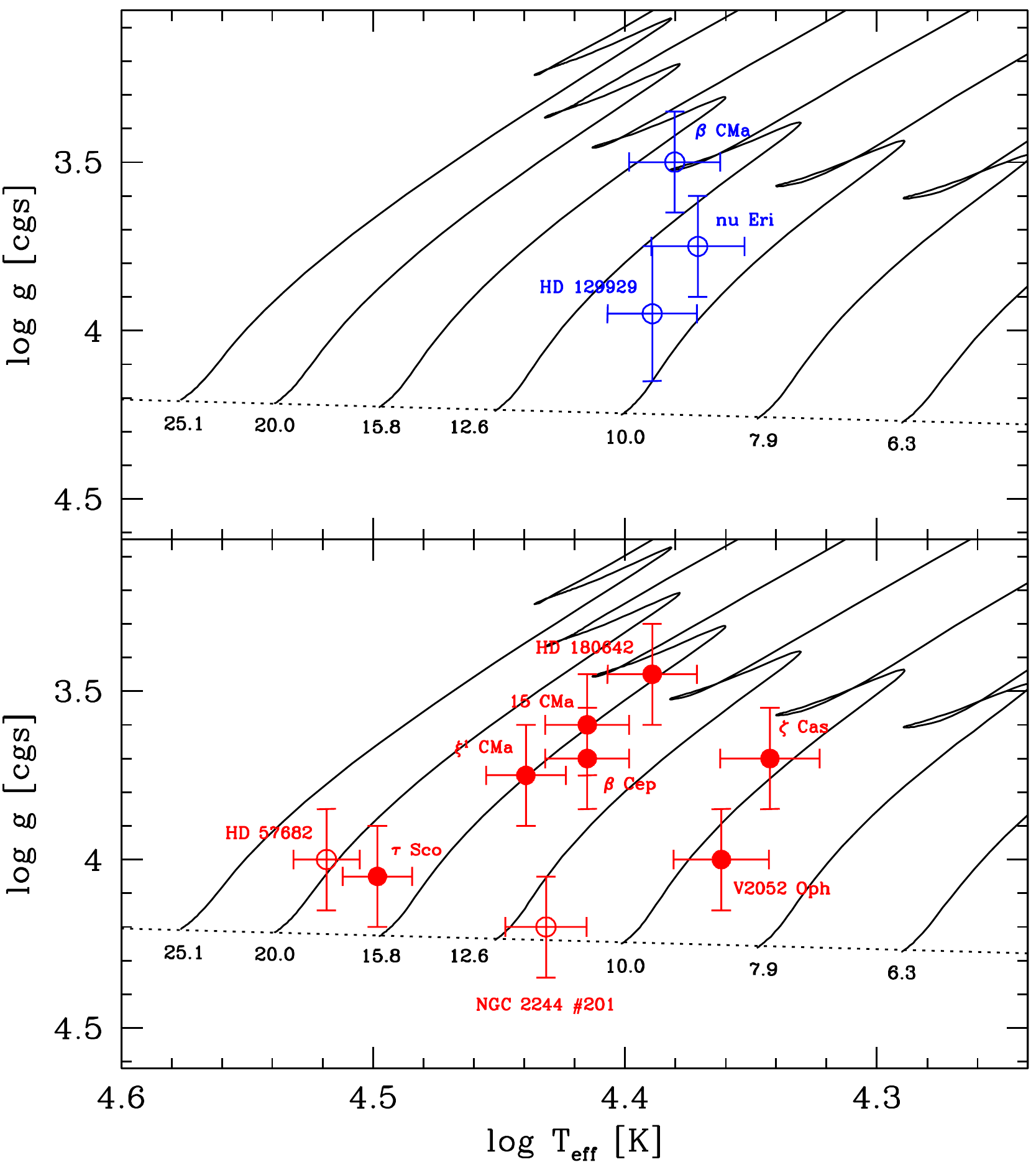}
\caption{Same as Fig.\ref{fig_teff_logg1}, but only taking into account the stars with a secure magnetic field (non) detection based on the spectropolarimetric observations of Alecian et al. (2008), Donati et al. (2001, 2006), Grunhut et al. (2009), Hubrig et al. (2009, 2010), Neiner et al. (2003a,b), Schnerr et al. (2006) and Silvester et al. (2009).\label{fig_teff_logg2}}
\end{minipage}
\end{figure}

However, it is likely that this simple relationship is only statistical and that other (still elusive) parameters may play a role in the amount of mixing experienced. This is particularly well illustrated by the cases of HD 57682 and $\tau$ Sco. These two stars have only slightly evolved off the ZAMS, have about the same mass (Fig.\ref{fig_teff_logg1}) and are slow rotators (likely for HD 57682 unless it is seen nearly pole on, while $\tau$ Sco has a true rotational velocity as low as 6 km s$^{-1}$; Donati et al. 2006). Yet, the latter exhibits evidence for CN-cycled material dredged up to the surface (an N excess of a factor $\sim$3; see also Przybilla, Nieva \& Butler 2008 who obtained [N/C]=--0.14$\pm$0.18), whereas the former does not. 

We conclude by noting that several aspects may complicate a direct comparison between the observed CNO surface abundances and the predictions of evolutionary models. First, the models including magnetic effects consider a field created through shear instabilities, whereas it has a simple morphology strongly indicative of a fossil origin in many stars. Second, many uncertainties about the evolution of the angular momentum along the main sequence still remain. For instance, while magnetic braking is unlikely to significantly spin down some of our magnetic targets (see, e.g., ud-Doula, Owocki \& Townsend 2009 in the case of $\zeta$ Cas), the impact of $g$-mode pulsations in redistributing the angular momentum in the slowly pulsating B stars (SPBs) or hybrid $\beta$ Cephei/SPBs in our sample might be dramatic (Townsend 2009). 

%
%
\section*{Acknowledgements}
The author acknowledges financial support from Belspo for contract PRODEX GAIA-DPAC. I am indebted to K. Butler for providing me with the line-formation codes DETAIL/SURFACE and S. Hubrig for having access to her unpublished magnetic data. Valuable suggestions from S. Sim\'on-D\'{\i}az (the referee), P. Williams (the editor), K. Pavlovski and E. Tamajo were also very much appreciated.

%
%
\footnotesize
\beginrefer

\refer Alecian E., Wade G. A., Catala C., et al., 2008, A\&A, 481, L99

\refer Asplund M., Grevesse N., Sauval A. J., Scott P., 2009, ARA\&A, 47, 481

\refer Brott I., Hunter I., de Koter A., Langer N., Lennon D., Dufton P., 2009, CoAst, 158, 55

\refer Butler K., Giddings J. R., 1985, in Newsletter of Analysis of Astronomical Spectra, No.9 (Univ. London)

\refer Claret A., 2004, A\&A, 424, 919

\refer Donati J.-F., Wade G. A., Babel J., Henrichs H. F., de Jong J. A., Harries T. J., 2001, MNRAS, 326, 1265

\refer Donati J.-F., Howarth I. D., Jardine M. M., et al., 2006, MNRAS, 370, 629

\refer Giddings J. R., 1981, Ph.D. Thesis, University of London

\refer Gies D. R., Lambert, D. L., 1992, ApJ, 387, 673

\refer Grunhut J. H., Wade G. A., Marcolino W. L. F., et al., 2009, MNRAS, 400, L94

\refer Grunhut J. H., Wade G. A., Marcolino W. L. F., et al., 2010, in proceedings of IAU Symp.272, `Active OB stars: structure, evolution, mass loss and critical limits', in press (arXiv:1009.3263)

\refer Halbedel E. M., 1993, IBVS, \#3850

\refer Heger A., Woosley S. E., Spruit H. C., 2005, ApJ, 626, 350

\refer Hillier D. J., Miller D. L., 1998, ApJ, 496, 407

\refer Hubrig S., Briquet M., Morel T., Sch\"oller M., Gonz\'alez J. F., De Cat P., 2008, A\&A, 488, 287

\refer Hubrig S., Briquet M., De Cat P., Sch\"oller M., Morel T., Ilyin I., 2009, AN, 330, 317

\refer Hubrig S., Ilyin I., Sch\"oller M., Briquet M., Morel T., De Cat P., 2010, ApJL, submitted

\refer Hunter I., Brott I., Lennon D. J., et al., 2008, ApJ, 676, L29

\refer Kilian J., 1992, A\&A, 262, 171

\refer Kilian J., 1994, A\&A, 282, 867

\refer Maeder A., Meynet G., 2005, A\&A, 440, 1041

\refer Morel T., Hubrig S., Briquet M., 2008, A\&A, 481, 453

\refer Morel T., Butler K., 2008, A\&A, 487, 307 

\refer Morel T., 2009, CoAst, 158, 122

\refer Neiner C., Geers V. C., Henrichs H. F., Floquet M., Fr\'emat Y., Hubert A.-M., Preuss O., Wiersema K., 2003a, A\&A, 406, 1019

\refer Neiner C., Henrichs H. F., Floquet M., et al., 2003b, A\&A, 411, 565

\refer Petit V., Wade G. A., Drissen L., Montmerle T., Alecian E., 2008, MNRAS, 387, L23

\refer Przybilla N., Nieva M.-F., Butler K., 2008, ApJ, 688, L103

\refer Schnerr R. S., Verdugo E., Henrichs H. F., Neiner C., 2006, A\&A, 452, 969

\refer Silvester J., Neiner C., Henrichs H. F., et al., 2009, MNRAS, 398, 1505

\refer Townsend R., 2009, in proceedings of `Stellar pulsation: challenges for theory and observation', AIPC, 1170, 355

\refer ud-Doula A., Owocki S. P., Townsend R. H. D., 2009, MNRAS, 392, 1022

\refer Vrancken M., Hensberge H., David M., Verschueren W., 1997, A\&A, 320, 878

\refer Wade G. A., Alecian E., Bohlender D. A., et al., 2010, in proceedings of IAU Symp.272, `Active OB stars: structure, evolution, mass loss and critical limits', in press (arXiv:1009.3563)

\endrefer           
\end{document}